# Nematic spin correlations in the tetragonal state of uniaxial strained BaFe$_{2-x}$Ni$_x$As$_2$

Xingye Lu,[1] J. T. Park,[2] Rui Zhang,[1] Huiqian Luo,[1] Andriy H. Nevidomskyy,[3] Qimiao Si,[3] and Pengcheng Dai[3,1,*]

[1]Beijing National Laboratory for Condensed Matter Physics, Institute of Physics, Chinese Academy of Sciences, Beijing 100190, China
[2]Heinz Maier-Leibnitz Zentrum (MLZ), Technische Universität München, D-85748 Garching, Germany
[3]Department of Physics and Astronomy, Rice University, Houston, Texas 77005, USA

**Understanding the microscopic origins of electronic phases in high-transition temperature (high-$T_c$) superconductors is important for elucidating the mechanism of superconductivity. In the paramagnetic tetragonal phase of BaFe$_{2-x}$T$_x$As$_2$ (where $T$ is Co or Ni) iron pnictides, an in-plane resistivity anisotropy has been observed. Here we use inelastic neutron scattering to show that low-energy spin excitations in these materials change from four-fold symmetric to two-fold symmetric at temperatures corresponding to the onset of the in-plane resistivity anisotropy. Because resistivity and spin excitation anisotropies both vanish near optimal superconductivity, we conclude that they are likely intimately connected.**

[*]To whom correspondence should be addressed. E-mail: pdai@rice.edu

Superconductivity in iron pnictides can be induced by electron or hole-doping of their antiferromagnetic (AF) parent compounds [1–6]. The parent compounds exhibit a tetragonal to orthorhombic structural phase transition at temperature $T_s$, followed by a paramagnetic to AF phase transition at $T_N$ ($T_s \geq T_N$) [4–6]. An in-plane resistivity anisotropy has been observed in uniaxially strained iron pnictides $BaFe_{2-x}T_xAs_2$ (where $T$ is Co or Ni) above $T_s$ [7–9]. This anisotropy vanishes near optimal superconductivity and has been suggested as a signature of the spin nematic phase that breaks the in-plane four-fold rotational symmetry ($C_4$) of the underlying tetragonal lattice [10-14]. However, such interpretation was put in doubt by recent scanning tunneling microscopy [15] and transport [16] measurements, which suggest that the resistivity anisotropy in Co-doped $BaFe_2As_2$ arises from Co-impurity scattering and is not an intrinsic property of these materials. On the other hand, angle-resolved photoemission spectroscopy (ARPES) measurements found that the onset of a splitting in energy between two orthogonal bands with dominant $d_{xz}$ and $d_{yz}$ character in the uniaxial strain detwinned samples at a temperature above $T_s$ [17, 18], thereby suggesting the involvement of the orbital channel in the nematic phase [19-22]. Here we use inelastic neutron scattering (INS) to show that low-energy spin excitations in $BaFe_{2-x}Ni_xAs_2$ (x = 0, 0.085 and 0.12) [23, 24] change from four-fold symmetric to two-fold symmetric in the uniaxial strained tetragonal phase at temperatures corresponding to the onset of the in-plane resistivity anisotropy.

The magnetic order of the parent compounds of iron pnictide superconductors is collinear, with the ordered moment aligned antiferromagnetically along the $a_o$-axis of the orthorhombic lattice (Fig. 1A), and occurs at a temperature just below $T_s \approx T_N \approx 138$ K for $BaFe_2As_2$ [5, 6]. Because of the twinning effect in the orthorhombic state, AF Bragg peaks from the twinned domains appear at the ($\pm 1$, 0) and (0,$\pm 1$) in-plane positions in reciprocal space (Fig. 1B) [3]. Therefore, one needs to prepare single domain samples by applying a uniaxial pressure (strain) along one-axis of the orthorhombic lattice to probe the intrinsic electronic properties of the system [7–9]. Indeed, transport measurements on uniaxial strain detwinned samples of electron-underdoped $BaFe_{2-x}T_xAs_2$ [7–10] reveal clear in-plane resistivity anisotropy even above the zero pressure $T_c$, $T_N$, and $T_s$ (Fig. 1C).

To search for a possible spin nematic phase [12–14], we carried out INS experiments in uniaxial strain detwinned parent compound $BaFe_2As_2$ ($T_N$ = 138 K), electron-underdoped superconducting $BaFe_{1.915}Ni_{0.085}As_2$ ($T_c$ = 16.5 K, $T_N$ = 44 K), and electron-overdoped superconducting $BaFe_{1.88}Ni_{0.12}As_2$ ($T_c$ = 18.6 K, tetragonal structure with no static AF order) [Fig. 1C] [23, 24] using a thermal triple-axis spectrometer. Horizontally and vertically curved pyrolytic graphite (PG) crystals were used as a monochromator and analyzer. To eliminate contamination from epithermal or higher-order neutrons, a sapphire filter was added before the monochromator and two PG filters were installed before the analyzer. All measurements were done with a fixed final wave vector, $k_f$ = 2.662 Å$^{-1}$. Our annealed square-shaped single crystals of $BaFe_2As_2$ (~120 mg), $BaFe_{1.915}Ni_{0.085}As_2$ (~220 mg), and $BaFe_{1.88}Ni_{0.12}As_2$ (~448 mg) were mounted inside aluminum based sample holders with an uniaxial pressure of $P \approx$ 15 MPa, ~7 MPa, and ~7 MPa, respectively, applied along the $a_o/b_o$ axes direction [25-27]. We define momentum transfer **Q** in three-dimensional reciprocal space in Å$^{-1}$ as $\mathbf{Q} = H\mathbf{a^*}+K\mathbf{b^*}+L\mathbf{c^*}$, where $H$, $K$, and $L$ are Miller Indies and $\mathbf{a^*} = \hat{\mathbf{a}}_\mathbf{o}2\pi/a_o$, $\mathbf{b^*} = \hat{\mathbf{b}}_\mathbf{o}2\pi/b_o$ and $\mathbf{c^*} = \hat{\mathbf{c}}2\pi/c$. In the AF ordered state of a 100% detwinned sample, the AF Bragg peaks should occur at ($\pm1,0,L$) ($L$ = 1, 3, 5, · · · ) positions in reciprocal space. In addition, the low-energy spin waves should only stem from the ($\pm1$, 0) positions with no signal at the (0,$\pm1$) positions [26,27]. By contrast, in the paramagnetic tetragonal phase ($T > T_s \geq T_N$) one would expect the spin excitations at the ($\pm 1, 0$) and ($0, \pm 1$) positions to have equal intensities [12,27].

The results of our INS experiments on uniaxial strain detwinned BaFe$_{2-x}$Ni$_x$As$_2$ are summarized in Figure 1C. The square red symbols indicate the temperature below which spin excitations at an energy transfer of $E = 6$ meV exhibit a difference in intensity between the ($\pm 1,0$) and ($0,\pm 1$) positions for undoped and electron underdoped BaFe$_{2-x}$Ni$_x$As$_2$. For electron overdoped BaFe$_{1.88}$Ni$_{0.12}$As$_2$, the same uniaxial pressure has no effect on spin excitations at wave vectors ($\pm 1,0$) and ($0,\pm 1$) [27]. A comparison to the transport measurements [10] in Fig. 1C indicates that the resistivity anisotropy occurs near the spin excitation anisotropy temperature $T^*$ determined from INS.

Given that our experiments are performed in uniaxial strain detwinned samples, it is important to establish how the structural and magnetic transition temperatures are affected by the applied pressure. Figure S2A compares the temperature dependence of the magnetic order parameters at (1,0,1)/(0,1,1) for BaFe$_2$As$_2$ in zero pressure (green symbols) and under uniaxial strain (red and blue symbols). We find that the BaFe$_2$As$_2$ sample is essentially 100% detwinned under the applied uniaxial strain without altering $T_N$ [27]. Similarly, the electron underdoped BaFe$_{1.915}$Ni$_{0.085}$As$_2$ is about 80% detwinned and has $T_N \approx 44$ K, unchanged from the zero-pressure case (fig. S2C) [27]. To investigate whether the tetragonal-to-orthorhombic structural phase transition in BaFe$_{2-x}$Ni$_x$As$_2$ is affected by uniaxial strain, we plot the temperature dependence of the (2,−2,0) nuclear Bragg peak of BaFe$_2$As$_2$; both zero pressure (fig. S2B) and detwinned samples (fig. S2D) exhibit a step-like feature at $T_s \approx 138$ K resulting from the vanishing neutron extinction effect due to the tetragonal-to-orthorhombic structural transition [28,29].

In previous spin-wave measurements on twinned BaFe$_2$As$_2$, a spin gap of ~10 meV was found at the (1,0,1) and (0, 1, 1) positions [30]. To probe spin excitations at the same wave vectors in the detwinned BaFe$_2$As$_2$, we aligned the sample in the [1,0,1]×[0,1,1] scattering plane [27]. Figure 2, A, C, and E, shows constant-energy scans centered at (1,0,1) approximately along the [1,$K$,1] direction. Whereas spin waves at (1,0,1) are clearly gapped at $E = 6$ meV in the AF ordered state ($T = 3$ K) in Fig. 2A, they are well-defined at $E = 15$ meV (Fig. 2C) and 19 meV (Fig. 2E), in line with the previous report [31]. We find no evidence for spin waves at $E = 6$, 15, and 19 meV at (0,1,1) (Fig. 2, B, D, and F, respectively), which is consistent with a nearly 100% detwinned BaFe$_2$As$_2$. On warming the system to the paramagnetic tetragonal state at $T = 154$ K, the spin gap disappears and the $E = 6$ meV spin excitations at the AF wave vector (1,0,1) are clearly stronger than those at (0, 1, 1) (Fig. 2, A and B) [27].

To quantitatively study the energy dependence of the spin excitation anisotropy in BaFe$_2$As$_2$ at a temperature above $T_s$, we plot in Fig. 3A the energy scans at wave vectors (1,0,1) and (0,1,1) and their corresponding backgrounds at $T = 154$ K [27]. The background subtracted scattering at (1,0,1) is consistently higher than that at (0,1,1) (Fig. 3C, and 3E, left inset). When we warm up to $T = 189$ K, the corresponding energy scans (Fig. 3B) and the signals above background (Fig. 3D) reveal that the differences at these two wave vectors disappear (Fig. 3E, left inset). Figure 3E shows the temperature dependence of the spin excitations (signal above background scattering) across $T_N$ and $T_s$. In the AF ordered state, we see only spin waves from the wave vector (1,0,1). On warming to the paramagnetic tetragonal state above $T_N$ and $T_s$, we see clear differences between (1,0,1) and (0,1,1) that vanish above ∼165 K, the same temperature below which anisotropy is observed in the in-plane resistivity (Fig. 3E, right inset) [32]. We conclude that the four-fold to two-fold symmetry change in spin excitations in $BaFe_2As_2$ occurs alongside the resistivity anisotropy.

To see if spin excitations in superconducting $BaFe_{1.915}Ni_{0.085}As_2$ also exhibit the four-fold to two-fold symmetry transition, we study the temperature dependence of the $E = 6$ meV spin excitations at the (1,0,1) and (0,1,1) wave vectors. In previous INS experiments on twinned $BaFe_{1.92}Ni_{0.08}As_2$, a neutron spin resonance was found near $E \approx 6$ meV [33]. Figure 4, A and C, shows approximate transverse and radial scans through (1,0,1) at various temperatures; one can clearly see the superconductivity-induced intensity enhancement from 48 K to 8 K. The corresponding scans through (0,1,1) (Fig. 4, B and D) have weaker intensity than those at (1,0,1). Figure 4E shows the temperature dependence of the magnetic scattering at (1,0,1) and (0,1,1). Consistent with constant-energy scans in Fig. 4, A-D, the scattering at (1,0,1) is considerably stronger than that at (0,1,1) above $T_c$. On warming through $T_N$ and $T_s$ [24], the spin excitation anisotropy between (1,0,1) and (0,1,1) becomes smaller, but reveals no dramatic change. The anisotropy disappears around $T^* \sim 80$ K, well above $T_N$ and $T_s$ (Fig. 4, E and F) but similar to the point of vanishing in-plane resistivity anisotropy [10]. Finally, we find that uniaxial strain does not break the $C_4$ rotational symmetry of the spin excitations in electron overdoped $BaFe_{1.88}Ni_{0.12}As_2$ [27]. In this compound, resistivity shows no $a_o/b_o$ anisotropy [10].

Conceptually, once the $C_4$ symmetry of the electronic ground state is broken, the electronic anisotropy will couple linearly to the orthorhombic lattice distortion $\epsilon = a_o - b_o$, so that the $C_4$ nematic transition should coincide with the tetragonal-to-orthorhombic transition at temperature $T_s$ [12–14]. How do we then understand the region $T_s < T < T^*$ in which the low-energy spin excitations develop an anisotropy? Theoretically, this is best understood in terms of the effective action for the electronic nematic order parameter $\Delta$ and magnetization $\mathbf{M}_{1/2}$ of the interpenetrating Néel sublattices [14,34,35]:

$$S_{eff}[\Delta, \mathbf{M}] = S_0[\mathbf{M}_1^2, \mathbf{M}_2^2] + \iint d\mathbf{q}\, d\omega [\alpha^{-1}(\mathbf{q}, \omega)|\Delta|^2 + v|\Delta|^4 - g(\mathbf{M}_1 \cdot \mathbf{M}_2)\Delta - \lambda \epsilon \Delta] + C_s \epsilon^2/2 \,,$$

(1)

Here $S_0$ is defined as part of the action that does not contain nematic correlations $(\mathbf{M}_1 \cdot \mathbf{M}_2)$ [36], which have been decoupled in terms of the bosonic field $\Delta(\mathbf{q}, \omega)$ characterized by the nematic susceptibility $\alpha(\mathbf{q}, \omega)$ where $E = \hbar\omega$, $g \approx 2$, $v$ is the quartic coupling among the bosonic fields, $\lambda$ is the linear coupling between the bosonic field and the orthorhombic lattice distortion $\epsilon$, and $\mathbf{q}$ is the momentum transfer within one Brillouin zone. Minimizing the action with respect to $\epsilon$, we arrive at $\epsilon = \lambda \langle \Delta \rangle / C_s$, where $C_s$ is the shear modulus. In other words, the orthorhombic lattice distortion is proportional to the nematic order parameter $\langle \Delta \rangle$ and both are expected to develop non-zero expectation values below $T_s$ [12-14]. However, the nematic field $\Delta$ undergoes fluctuations in the tetragonal phase above $T_s$ while $\epsilon$ remains zero. These fluctuations will be observable in dynamic quantities, such as the finite-energy spin fluctuations, and in transport measurements. We therefore conclude that the scale $T^*$, below which we observe anisotropy of low-energy spin fluctuations (Figs. 3E, 4E, and 4F) and where the resistivity anisotropy is observed (Fig. 3E, right inset) marks a typical range of the nematic fluctuations.

Several remarks are in order. First, the applied uniaxial pressure used to detwin the samples will induce a finite value of $\epsilon$ at any temperature, so that strictly speaking, the structural transition at $T_s$ will be rendered a crossover. In practice, however, the applied pressure is too small to cause a perceptible lattice distortion, which is why the transition temperature $T_s$, as determined from the extinction effect of the nuclear (2,−2,0) Bragg peak remains unchanged from the zero pressure case [fig. S2B and S2D] [26,27]. On the other hand, the extent of nematic fluctuations may be sensitive to the shear strain, in agreement with the reported increase of $T^*$ (as determined from resistivity anisotropy) with the uniaxial pressure [37]. Second, in Eq. 1 the variable $\Delta$ could equally signify the orbital order $\Delta \propto (n_{xz} - n_{yz})$ which lifts the degeneracy between the Fe $d_{xz}$ and $d_{yz}$ orbitals. In fact, the two order parameters will couple linearly to each other, $(\mathbf{M}_1 \cdot \mathbf{M}_2) \propto (n_{xz} - n_{yz})$, so that both will develop a non-zero value below $T_s$. In this respect, our findings are also consistent with the recent ARPES finding of an orbital ordering [17,18] in BaFe$_2$As$_2$. This underlines the complementarity of the spin-nematic and orbital descriptions of the $C_4$ symmetry breaking. Third, in the nearly optimally electron-doped superconductor, we observe anisotropy of the low-lying spin excitations in the tetragonal phase $T_s < T < T^*$ even though the orbital order is no longer detectable by ARPES [17, 18]. This is consistent with the absence of a static nematic order $\langle \Delta \rangle = 0$ above $T_s$, whereas the observed spin anisotropy originates from Ising-nematic fluctuations. Because $T_s$ is considerably suppressed for this doping, these fluctuations are quantum rather than thermal: They persist beyond the immediate vicinity of $T_s$, and the associated spin anisotropy should have sizable dependence on frequency that can be probed by future experiments. Fourth, when resistivity anisotropy under uniaxial strain disappears in the electron-overdoped sample [10], the uniaxial-strain-induced spin excitation anisotropy also vanishes [Fig. 1E and fig. S5], which suggests a direct connection between these two phenomena. Finally, our measurements in the spin channel do not necessarily signal a thermodynamic order at the temperature $T^*$. Rather, $T^*$ likely signals a crossover, whereas the true nematic transition occurs at $T_s$ [9]. This implies that a static order above $T_s$ inferred from recent measurements of magnetic torque anisotropy in the isovalent $BaFe_2As_{2-x}P_x$ [38] is most likely not in the spin channel accessible to the inelastic neutron scattering. A static order in other channels—, for instance, an octupolar order,—would, however, not contradict our observations.

We thank Wenliang Zhang for help during the experiments. The work at the Institute of Physics, Chinese Academy of Sciences is supported by Ministry of Science and Technology of China (973 project: 2012CB821400 and 2011CBA00110), National Natural Science Foundation of China and China Academy of Engineering Physics. The work at Rice is supported by the U.S. NSF-DMR-1308603 and DMR-1362219 (P.D.), by Robert A. Welch Foundation grant no. C-1839 (P.D.), C-1818 (A.H.N.), and no. C-1411 (Q.S.), and by the U.S. NSF-DMR-1309531 and the Alexander von Humboldt Foundation (Q.S.). A.H.N. and Q.S. acknowledge the hospitality of the Aspen Center for Physics, where support was provided by NSF grant PHYS-1066293.

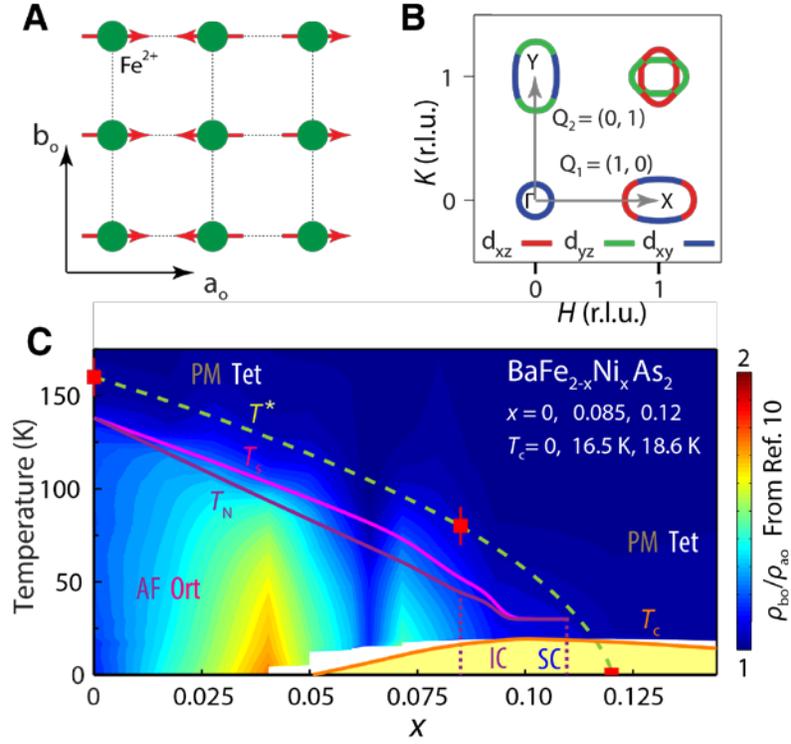

FIG. 1: **Summary of the results**. (**A**) The AF spin arrangement of iron in the FeAs layer of BaFe$_2$As$_2$. (**B**) The corresponding Fermi surfaces with one circular hole pocket around the zone center Γ point and two elliptical electron pockets at X and Y points [3]. (**C**) The electronic phase diagram of BaFe$_{2-x}$Ni$_x$As$_2$ from resistivity anisotropy ratio $\rho_{bo}/\rho_{ao}$ obtained under uniaxial pressure [10]. The spin excitation anisotropy temperatures are marked as $T^*$. The AF orthorhombic (Ort), incommensurate AF (IC) [23], paramagnetic tetragonal (PM Tet), and superconductivity (SC) phases are marked.

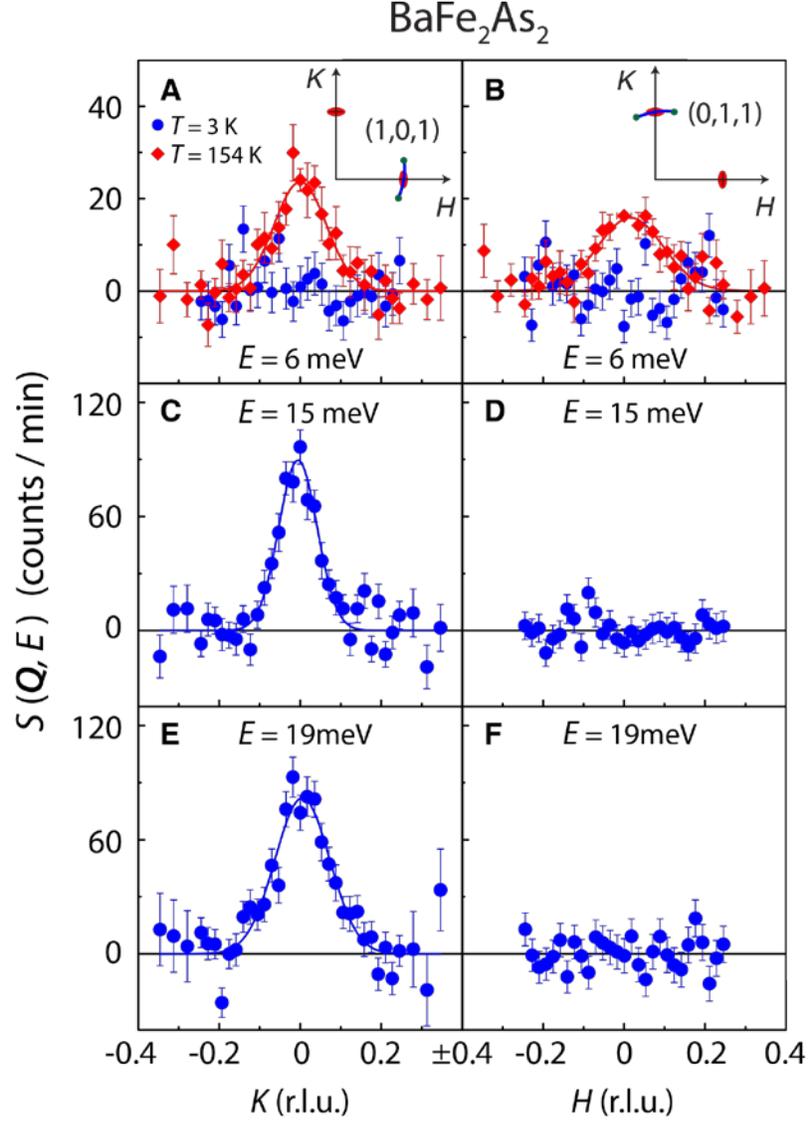

FIG. 2: **Constant-energy scans for detwinned BaFe$_2$As$_2$.** The $E = 6$ meV rocking scans measured at $T = 3$ K in the AF ordered state and $T = 154$ K in the paramagnetic tetragonal state centered at **(A)** (1,0,1) and **(B)** (0,1,1). $T = 3$ K scans at $E = 15$ meV for **(C)** (1,0,1) and **(D)** (0,1,1) and at $E = 19$ meV for **(E)** (1,0,1) and **(F)** (0,1,1). The in-plane projected trajectories of the rocking scans crossing (1,0,1) and (0,1,1) are illustrated by blue lines in the insets of (A) and (B), respectively. Solid lines are Gaussian fits.

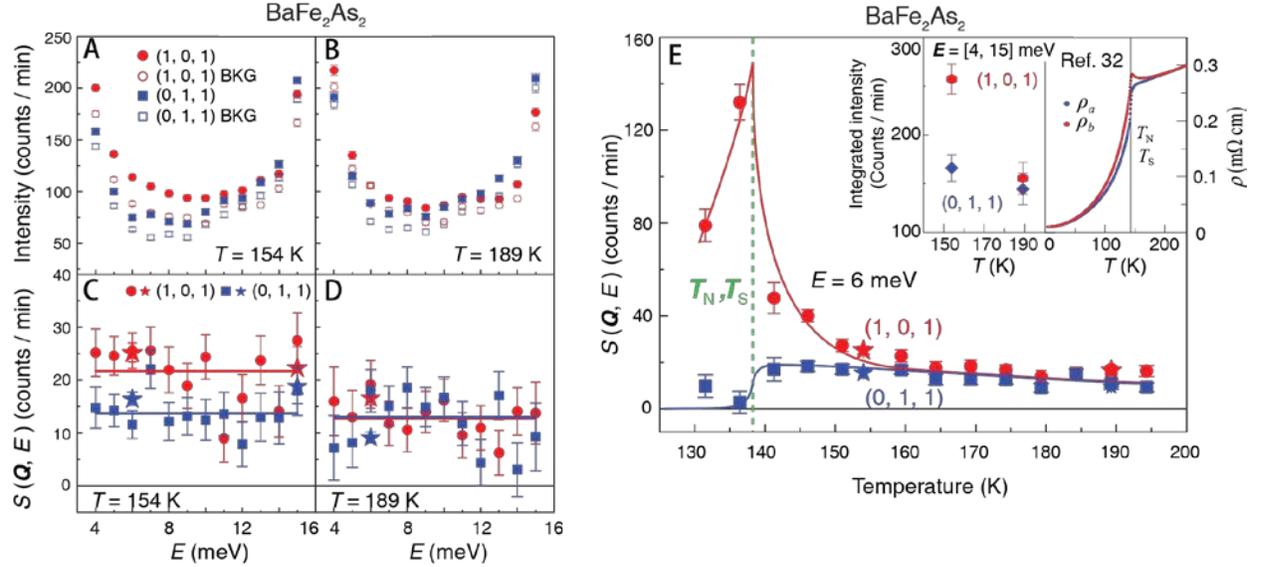

FIG. 3: **Temperature dependence of spin excitations for BaFe$_2$As$_2$.** Energy scans at wave vectors (1,0,1)/(0,1,1) and corresponding background positions at temperatures above $T_s$: **(A)** $T =$ 154 K and **(B)** 189 K. **(C and D)** Magnetic scattering after subtracting the backgrounds. The solid lines are constant line fits. **(E)** Temperature dependence of the spin excitations at $E = 6$ meV for (1,0,1) and (0,1,1). The anisotropy in spin excitations vanishes around $T = 160 \pm 10$ K. The data marked by filled squares and dots in (C) to (E) were obtained by subtracting the corresponding backgrounds. For each temperature, the background intensities at $\mathbf{Q} = (1,0,1)$ and (0,1,1) were obtained by averaging the data at the two wave vectors marked by green dots in the insets of Fig. 2, A and B, respectively. The data denoted by filled stars in (C) to (E) were obtained by fitting the rocking scans for $E = 6, 15$ meV at different temperatures. The solid lines in (E) are guides to the eye. The left inset in (E) shows temperature dependence of the integrated intensity from 4 to 15 K, the right inset shows temperature dependence of the resistivity from [32].

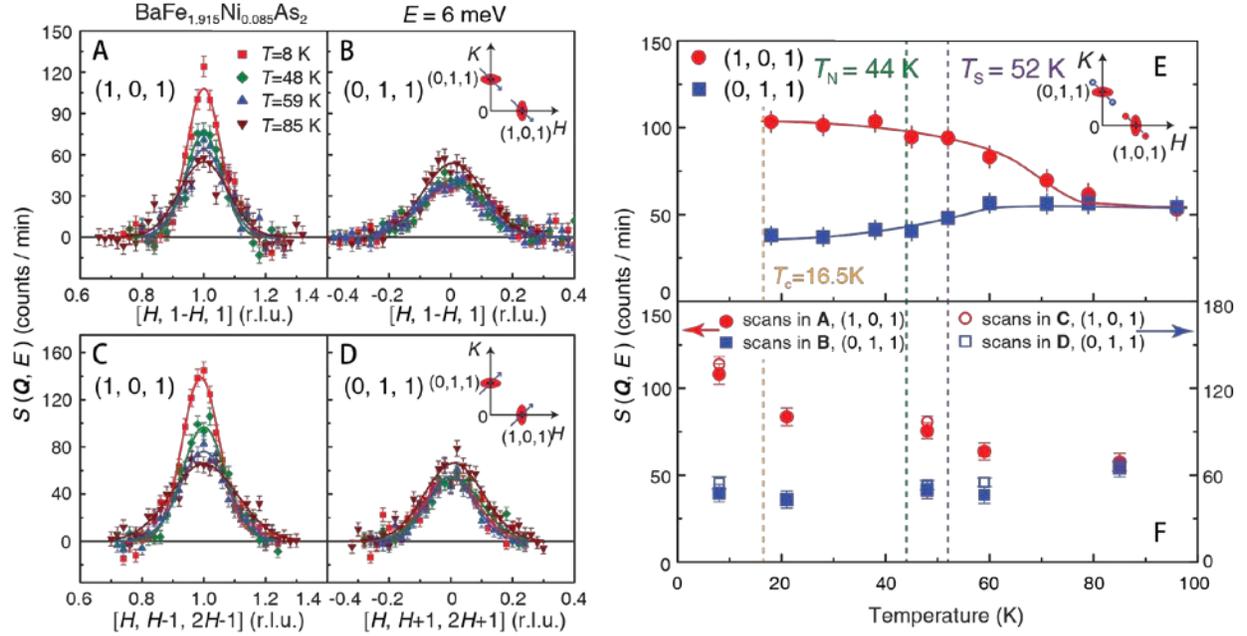

FIG. 4: **Temperature dependence of spin excitations for BaFe$_{1.915}$Ni$_{0.085}$As$_2$**. Background subtracted **Q**-scans at $E = 6$ meV around **(A, C)** (1, 0, 1) and **(B, D)** (0, 1, 1). The trajectory of the scans (blue line) crossing spin excitations (red ellipses) are illustrated in the insets of (B) and (D). **(E, F)** Temperature dependence of the spin excitations at (1,0,1) and (0,1,1) at $E = 6$ meV. The data in (E) were obtained by subtracting the background intensity from the peak intensity at every temperature; the data in (F) were obtained by fitting the **Q** scans. The wavevectors used for calculating the background (blue circles and red dots) are shown in the inset of (E). The solid curves in (A-D) are Gaussian fits and in (E) are guides to the eye. The temperatures for structural (purple), magnetic (blue), and SC (orange) transitions are marked by vertical dashed lines [24].

# Supplementary Materials for

## Nematic Spin Correlations in the Tetragonal State of Uniaxial Strained $BaFe_{2-x}Ni_xAs_2$

Xingye Lu, J. T. Park, Rui Zhang, Huiqian Luo, Andriy H. Nevidomskyy, Qimiao Si, and Pengcheng Dai*

*To whom correspondence should be addressed. Email: pdai@rice.edu

**This PDF file includes:**

Materials and Methods
Figs. S1 to S5
References

**Materials and Methods**

**Sample preparation**

BaFe$_{2-x}$Ni$_x$As$_2$ single crystals were grown with self-flux method. The basic sample characterizations were described in our previous study [39]. Large single crystals with less flux were selected and annealed in Ba$_2$As$_3$ for several days. The tetragonal [1,1,0] direction of annealed crystals was determined by X-ray Laue diffraction. The crystals were cut into rectangular pieces along the [1,1,0] and [1,-1,0] directions by high precision wire saw.

Samples used in this report are BaFe$_2$As$_2$ ($T_N = T_s = 138$ K), electron-underdoped superconducting BaFe$_{1.915}$Ni$_{0.085}$As$_2$ ($T_c = 16.5$K, $T_N = 44$ K, $T_s = 52$ K) and electron-overdoped BaFe$_{1.88}$Ni$_{0.12}$As$_2$ ($T_c = 18.6$ K).

**Device for sample detwinning**

The device for sample detwinning was made of 6061 aluminum alloy with low neutron incoherent scattering cross section. As shown in Fig. S1A, uniaxial pressure can be applied by a spring along orthorhombic [0,1,0] direction by tuning the screw in one end. The pressure can be calculated by the known elasticity coefficient ($k = 10.5$ N/mm) and the compression of the spring ($\Delta x$). The elasticity coefficient was measured in our lab, as shown in Fig. S1B. Take BaFe$_{1.915}$Ni$_{0.085}$As$_2$ as an example, the sample size is 7.97mm*6.42mm*0.70mm. Thus the sectional area we applied pressure were 7.97*0.7=5.58 mm$^2$. The compression of the spring in the experiment was about 3.5 mm. A simple calculation gives a pressure ~6.6 MPa. However, since the pressure calibrations were done at room temperature and we do not know the elasticity coefficient of the spring at low temperatures, the applied pressure was only a rough estimate. What we do know is that the applied pressure is sufficient to detwin the sample. In Fig. S1A, the device is mounted on a supporting sample holder to align the crystal in the [1,0,1]×[0,1,1] scattering plane (see Fig. S1C), where the spin excitations at $\mathbf{Q}_{AF} = (1,0,1)$ and (0,1,1) can be measured and compared directly.

## Sample detwinning

Our neutron scattering experiments were carried out using the PUMA thermal triple-axis spectrometer at the MLZ in Garching, Germany. The efficiency of sample detwinning can be checked by comparing the intensity of the magnetic Bragg peaks at (1,0,1) and (0,1,1). For a fully detwinned sample, magnetic elastic scattering is expected to appear only at the antiferromagnetic wave vector (1,0,$L$=1, 3, ···). Figure S1D shows the measurements of magnetic order at the (1,0,1) and (0,1,1) positions. A magnetic Bragg peak was observed at the (1,0,1) position while no magnetic signal was detected at (0,1,1), indicating that the $BaFe_2As_2$ single crystal was completely detwinned. Above the antiferromagnetic transition temperature ($T_N$), the magnetic peak at (1,0,1) disappears as expected. The temperature dependence of the magnetic Bragg peak at (1,0,1) and (0,1,1) was also measured for $BaFe_2As_2$ and $BaFe_{1.915}Ni_{0.085}As_2$ (Figs. S2, A, B). Consistent with magnetic Bragg peak measurements in Fig. S1D, the $BaFe_2As_2$ was fully detwinned with most of the magnetic intensity located at (1,0,1). For the $BaFe_{1.915}Ni_{0.085}As_2$, the sample was partially detwinned with the intensity at (1,0,1) about three times larger than that at (0,1,1). In both cases, the applied pressure did not change $T_N$. We have also measured the effect of uniaxial pressure on structural transition of $BaFe_2As_2$. As shown in Figs. S2, C and D, the temperature dependence of the intensity at the (2,-2,0) nuclear Bragg reflection for the twinned and detwinned samples both shows a dramatic jump at $T_s = 138$ K arises from the neutron extinction release that occurs due to strain and domain formation related to the orthorhombic distortion, indicating that the uniaxial pressure does not change the tetragonal to orthorhombic lattice transition temperature [28,29]. In previous work [29], the measurable extinction release at temperatures well above $T_s$ was suggested as arising from the significant structural fluctuations related to the orthorhombic distortion. If we assume this interpretation is correct, our data for the detwinned sample would suggest that uniaxial pressure pushes structural fluctuations to a temperature similar to resistivity anisotropy. Since extinction effect is typically only found for strong nuclear Bragg peaks, we do not expect to observe similar effect in weak magnetic Bragg scattering in Fig. S2A.

We also note that the (1,0,1) magnetic Bragg peak intensity in the twinned sample (Fig. S2A, green symbols) is not 1/2 of the detwinned sample. Since we must take the sample outside the cryostat to release the uniaxial pressure, the remounted sample will likely be at a slightly different location inside the neutron beam. Thus, for sharp magnetic or nuclear Bragg peaks, the measured intensity for strained and unstrained samples can only be compared approximately. However, this does not affect our experimental conclusion since the comparison of the scattering intensities at two wave vectors $\mathbf{Q}_{AF}$=(1,0,1) and (0,1,1) and their temperature dependence was done under the identical setup for each pressed or ambient-pressure case.

As the BaFe$_2$As$_2$ crystal was detwinned by uniaxial pressure, the low-energy spin waves were also fully "detwinned". In a twinned sample, the spin waves show four-fold symmetry owing to the existence of twin domains as seen from our neutron time-of-flight measurements for BaFe$_2$As$_2$ (Fig. S3A) [31]. On warming to above $T_s=T_N$, spin excitations still obey the four-fold rotational symmetry (Fig. S3C). For a uniaxial strained BaFe$_2$As$_2$, on the other hand, the spin waves stem only from the antiferromagnetic wave vector $\mathbf{Q}_{AF} = (1,0,L)$, as shown in Fig. S3B. In the tetragonal state ($T > T_s = T_N$), the magnetic order disappears and the paramagnetic spin excitations of the unconstrained sample show four-fold symmetry (Fig. S3C).

**Background subtraction**

As in typical magnetic neutron scattering experiments, the backgrounds of constant-energy scans are temperature dependent. Raw inelastic neutron scattering data for typical constant-energy scans are shown in Fig. S4A. The solid curves in Fig. S4A are single Gaussian fits assuming a linear background. Apparently, most of the scans in our measurements are well described by a Gaussian with a linear background. Figure S4B shows the linear-background subtracted data from raw data in Fig. S4A.

**Results of BaFe$_{1.88}$Ni$_{0.12}$As$_2$**

Different from the parent compound (x = 0) and under-doped (x=0.085) sample, the uniaxial pressure has no effect on the $C_4$ rotational symmetry of the spin excitations for slightly electron over-doped BaFe$_{1.88}$Ni$_{0.12}$As$_2$ sample [40]. Figures S5, A and B show constant-energy scans at $E = 6$ meV below and above $T_c$, respectively, under the same uniaxial strain as that of the superconducting BaFe$_{1.915}$Ni$_{0.085}$As$_2$. We find that spin excitations have identical intensity at the (1,0,1) and (0,1,1) wave vectors, thus preserving the $C_4$ rotational symmetry of the underlying lattice. Figures S5, C and D show constant-momentum scans across $T_c$ at (1,0,1) and (0,1,1), respectively. The temperature difference plot confirms the neutron spin resonance with identical intensity and energy ($E = 7$ meV) for (1,0,1) and (0, 1, 1) wave vectors [40]. Figure S5F plots the temperature dependence of the scattering intensity at $E = 7$ meV which proves that the applied uniaxial strain does not break the $C_4$ rotational symmetry of the spin excitations in the electron overdoped compound.

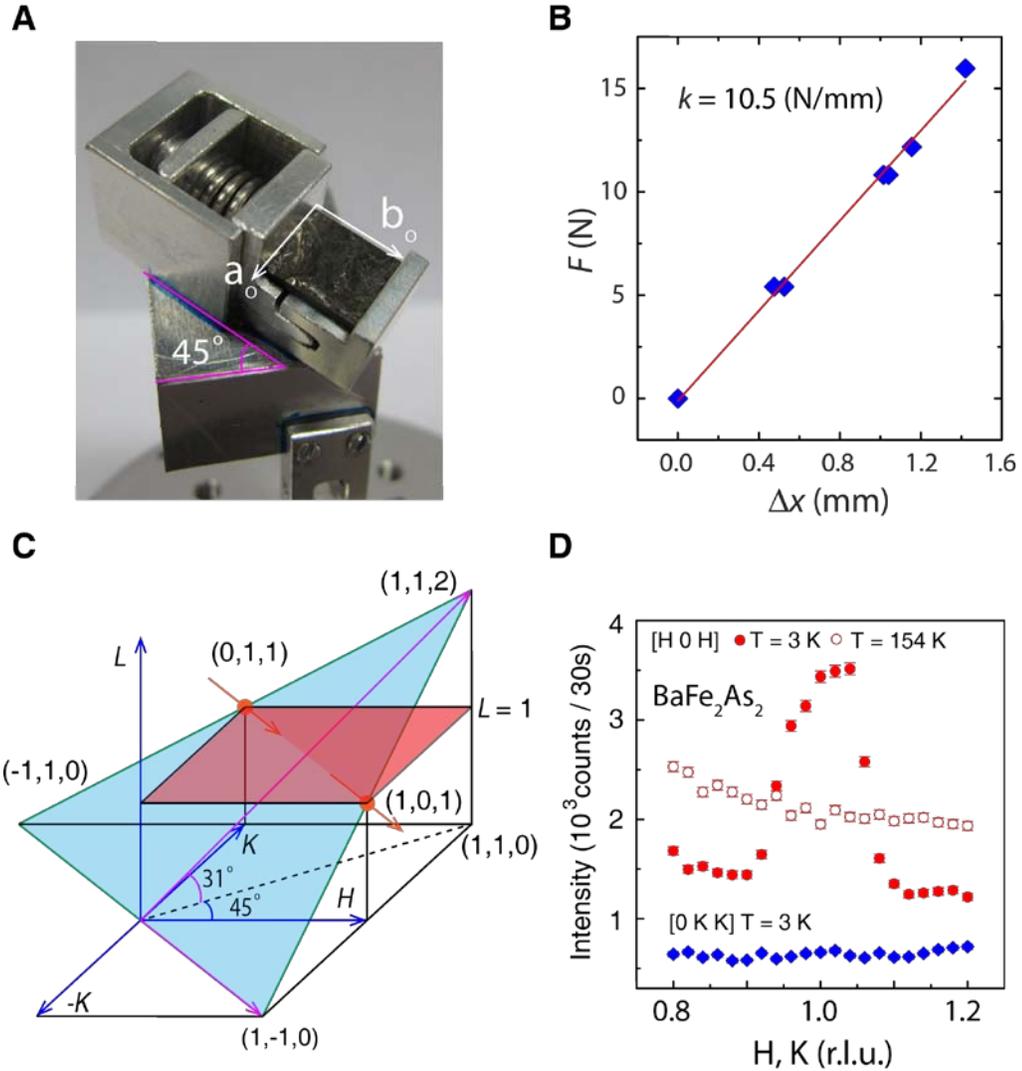

**Figure S1**: (**A**) Detwinning device mounted on the supporting sample holder. The crystal was aligned in the [1,0,1]×[0,1,1] scattering plane. Uniaxial pressure can be applied by a steel spring driven by a screw. (**B**) Measurement of elasticity coefficient of the spring. (**C**) The light blue shaded zone is the scattering plane where $\mathbf{Q}_{AF} = (1,0,1)$ and $(0,1,1)$ can be measured and compared directly. The (1,0,1) and (0,1,1) positions are marked by orange dots. (**D**) Magnetic Bragg peaks at the (1,0,1) and (0,1,1) positions in the detwinned $BaFe_2As_2$ below and above $T_N$.

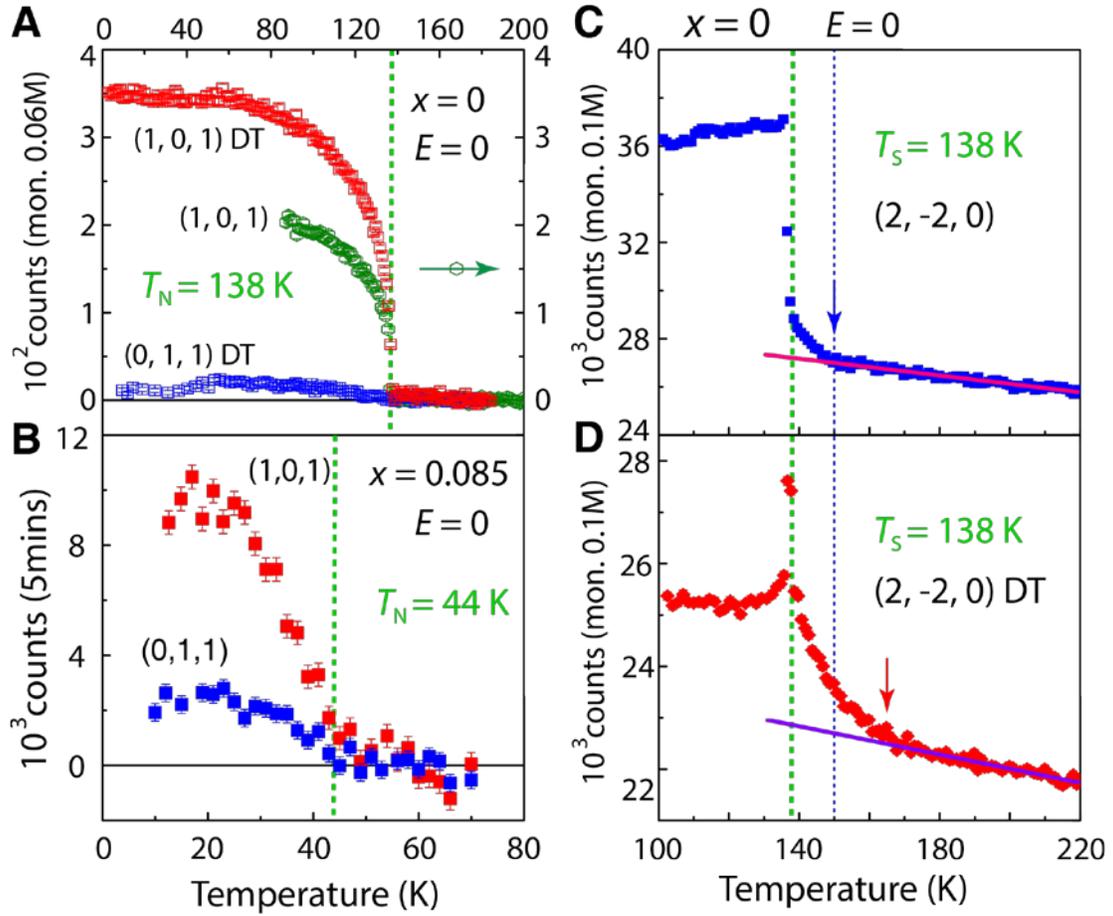

**Figure S2**: (**A**) Magnetic Bragg peak intensity at the (1,0,1) and (0,1,1) positions for the BaFe$_2$As$_2$ at zero pressure (green) and $P \sim 15$ MPa uniaxial pressure along the $b_o$ axis. (**B**) Similar Bragg peaks for BaFe$_{1.915}$Ni$_{0.085}$As$_2$. (**C**) Temperature dependence of the (2,-2,0) nuclear Bragg peak at zero pressure. The sharp step at Ts is caused by releasing of the neutron extinction due to tetragonal to orthorhombic lattice distortion. (**D**) The identical scan under $P \sim 15$ MPa uniaxial pressure.

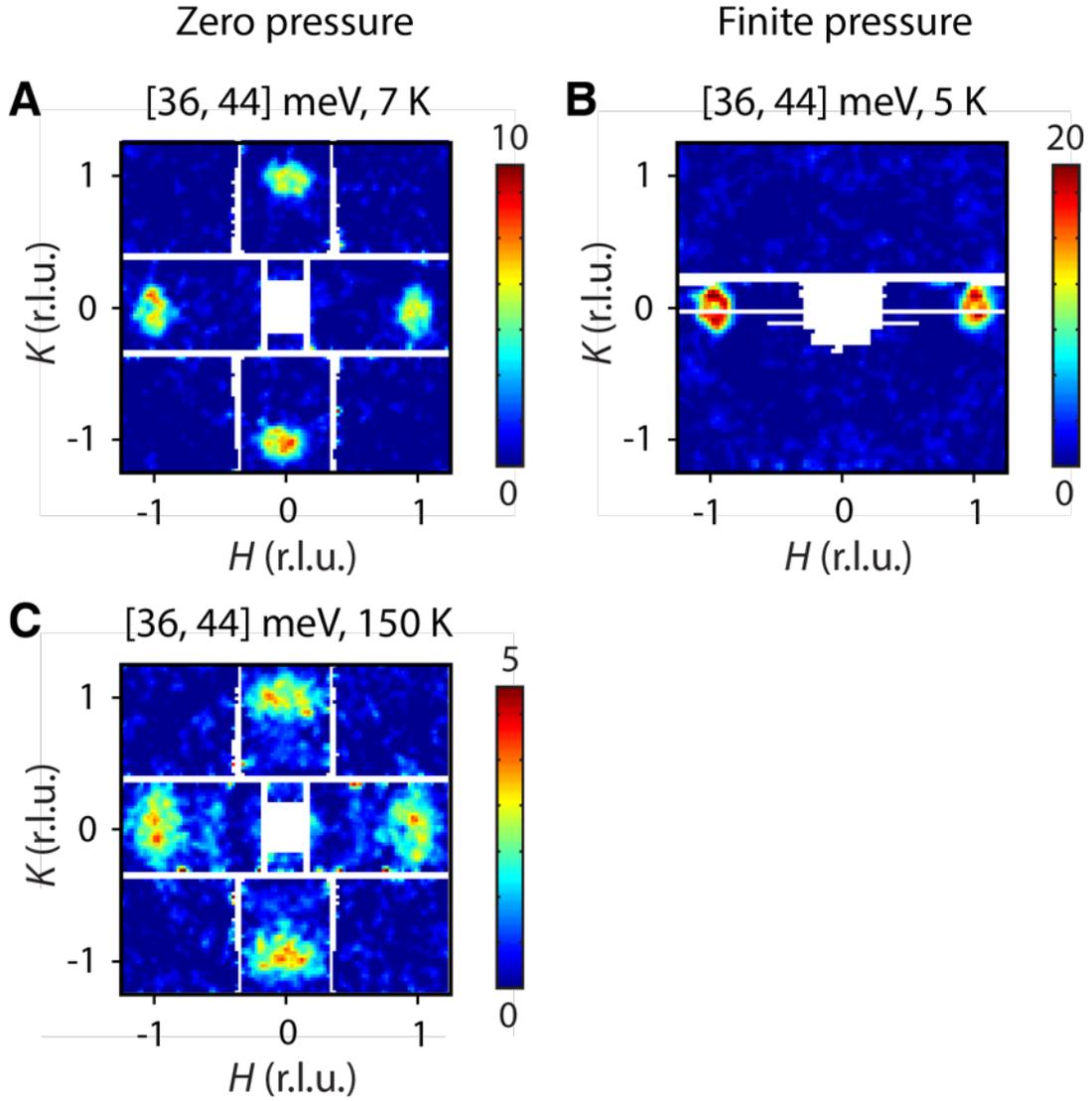

**Figure S3**: Comparison of the low energy spin waves for the twined and nearly 100% detwinned $BaFe_2As_2$. (**A**) Constant energy slice of the low-energy spin waves for the twinned $BaFe_2As_2$. Data were collected on MAPS time-of-flight spectrometer with $E_i = 80$ meV at $T = 7$ K [31]. The spin waves show four-fold symmetry due to twinning. (**B**) Identical slice of the detwinned $BaFe_2As_2$ collected on MERLIN time-of-flight spectrometer with $E_i = 80$ meV at $T = 5$ K (unpublished data). Detwinning transfers the magnetic intensity from (0,1,1) to (1,0,1). (**C**) At temperature $T = 150$ K $> T_N$, the paramagnetic spin excitations of the twinned sample show four-fold symmetry, consistent with the four-fold symmetry of the underlying tetragonal lattice.

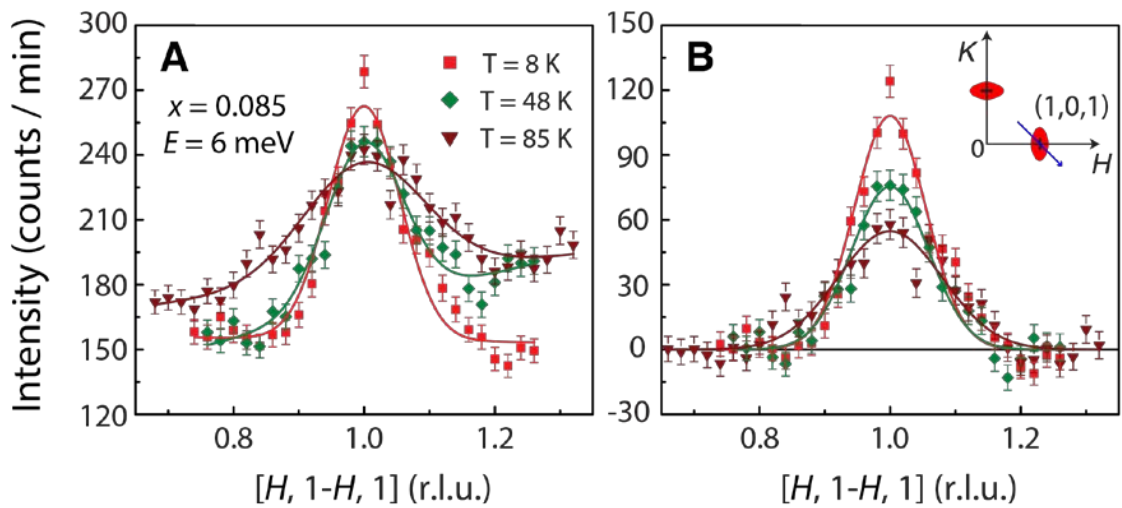

**Figure S4**: Background subtraction of inelastic neutron scattering data. (**A**) The raw data of the transverse scan across the (1,0,1) position. The scan trajectory is shown in the inset of (**B**). The solid curves are Gaussian fits of the data on linear backgrounds. (**B**) The linear-background subtracted data.

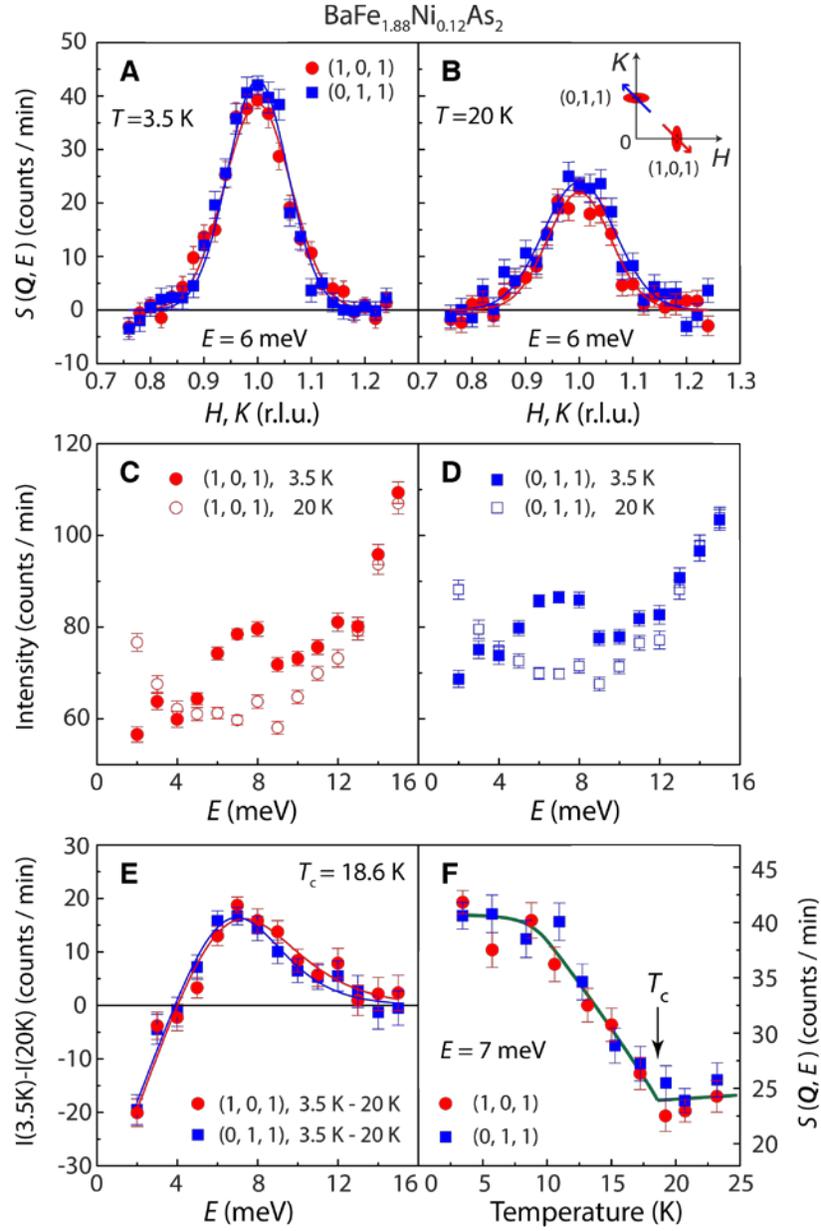

**Figure S5**: Temperature dependence of spin excitations at (1,0,1) and (0,1,1) for BaFe$_{1.88}$Ni$_{0.12}$As$_2$ [40]. (**A**) Linear background subtracted **Q**-scans around (1,0,1) and (0,1,1) at $E = 6$ meV and 3.5 K. (**B**) Identical scans at 20 K. The inset shows the trajectory of the scans. Constant **Q**-scans at (**C**) (1,0,1) and (**D**) (0,1,1) below and above $T_c$. (**E**) Temperature difference plot showing neutron spin resonance at (1,0,1) and (0,1,1). (**F**) Temperature dependence of the scattering at the resonance energy at (1,0,1) and (0,1,1).